\documentclass[12pt]{iopart}

\usepackage{amsfonts}
\input epsf

\def\bce{\begin{center}}
\def\ece{\end{center}}
\def\beq{\begin{eqnarray}}
\def\eeq{\end{eqnarray}}
\def\ben{\begin{enumerate}}
\def\een{\end{enumerate}}
\def\bei{\begin{itemize}}
\def\eei{\end{itemize}}

\def\ni{\noindent}
\def\nn{\nonumber}

\def\ms{\medskip}

\def\brr{\begin{array}}
\def\err{\end{array}}

\renewcommand{\Re}{\mbox{\rm Re}}

\newcommand{\w}{\omega}
\newcommand{\f}{\frac}

\renewcommand{\tilde}{\widetilde}

\begin{document}

\title[]{Dynamical Casimir Effect with Semi-Transparent Mirrors, and Cosmology\footnote{Talk given in the
Workshop ``Quantum Field Theory under the Influence of External
Conditions (QFEXT07)'', Leipzig (Germany), September 17 - 21, 2007}}

\author{Emilio Elizalde}

\address{Instituto de Ciencias del Espacio (CSIC) \\
Institut d'Estudis Espacials de Catalunya (IEEC/CSIC) \\
Campus UAB, Facultat de Ci\`encies, Torre C5-Parell-2a planta \\
E-08193 Bellaterra (Barcelona) Spain} \ead{elizalde@ieec.uab.es
http://www.ieec.fcr.es/english/recerca/ftc/eli/eli.htm}

\begin{abstract}
After reviewing some essential features of the Casimir effect and,
specifically, of its regularization by zeta function and Hadamard
methods, we consider the dynamical Casimir effect (or Fulling-Davis
theory), where related regularization problems appear, with a view
to an experimental verification of this theory. We finish with a
discussion of the possible contribution of vacuum fluctuations to
dark energy, in a Casimir like fashion, that might involve the
dynamical version.

\end{abstract}

\maketitle

\section{Introduction}
It was observed some time ago that the universe expansion
accelerates. This important discovery is still in search for an
explanation. It could in fact be found within Einsteinian gravity,
even if the only possibility there seems to be to consider the
contributions of the quantum vacuum fluctuations of the fields
pervading the universe to the cosmological constant, as was
discussed by Zel'dovich in a quite convincing way many years before
the acceleration of that expansion was discovered \cite{zeld1}. This
would be nice and, in principle, requires no new physics, however
there are several problems, as: (i) the {\it cosmological constant
problem}, that is, the contribution of the vacuum fluctuations seems
to be exceedingly large, as compared with the mentioned
astrophysical observations, and (ii) the {\it coincidence problem},
related with the fact that in relative terms the associated energy
is, in the present epoch, such a large part (over 72\%) of the whole
energy content of the universe (that is, of the same order of
magnitude and even dominating the energy content of the universe. If
we pay the price to modify Einstein's theory, then things become
easier to adjust, but other problems emerge. One cannot be happy
with such a number of possibilities, with increasing number of
parameters, and many are just effective or phenomenological models:
tensor, scalar-tensor, phantom, etc.

Here we will discuss a couple of specific problems of the vacuum
fluctuations approach only, some of rather technical, other of more
fundamental nature, in relation with regularization of quantum field theories in the
presence of boundaries and with the so-called {\it dynamical}
Casimir effect. We will recall a piece of sound mathematics needed
for the regularization issue. Then we will explicitly illustrate an
important aspect of this issue, namely the introduction of
physically meaningful regularization parameters, for the case of the
dynamical Casimir effect. We will finish with a discussion of
possible cosmological imprints of the Casimir effect, in some
particular models.

\section{Casimir Effect: On the Boundary Divergences}
Imposing mathematical boundary conditions on physical quantum
fields is not easy, as first discussed  by Deutsch and Candelas
\cite{dc79}, who quantized em and scalar fields near a smooth
boundary and calculated the renormalized vacuum expectation value of
the stress-energy tensor, to find that the energy density {\it
diverges} as the boundary is approached. Regularization and
renormalization did not seem to cure the problem with infinities in
this case and an infinite {\it physical} energy was obtained if the
mathematical boundary conditions were to be fulfilled. In an attempt to solve this,
the authors argued that physical surfaces have nonzero depth, and
this could be taken as a dimensional cutoff to regularize the
infinities. Later, Kurt Symanzik did a rigorous analysis of quantum field theory in
the presence of boundaries \cite{ks81}. Prescribing the value of the
quantum field on a boundary means using the Schr\"odinger
representation, and Symanzik was able to show it to exist to all
orders in the perturbative expansion. The issue was proven to be
meaningful within the domains of renormalized quantum field theory. In this case the
boundary conditions and the hypersurfaces themselves were treated at a pure
mathematical level (zero depth) by using delta functions.

New approaches to the problem have been postulated recently (see
e.g. \cite{bj1}). boundary conditions on a field, $\phi$, are enforced on a surface,
$S$, by introducing a scalar potential, $\sigma$, of Gaussian shape
living on and near the surface. When it becomes a delta function,
the boundary conditions (Dirichlet here) are enforced: the delta-shaped potential
kills {\it all} the modes of $\phi$ at the surface. For the rest,
the quantum system undergoes a full-fledged quantum field theory renormalization, as
in the case of Symanzik's approach. The results obtained confirm
\cite{dc79} in the several models studied but do not agree with
\cite{ks81}. They are also in contradiction with many textbooks and
review articles dealing with the Casimir effect \cite{cb1}, where no
infinite energy density when approaching the Casimir plates had been
reported.

In some circumstances specific regularization methods have been
employed with success, as zeta function \cite{zetabas1} and Hadamard
regularization, this last in higher-post-Newtonian general
relativity \cite{had1} and in recent variants of axiomatic and
constructive quantum field theory \cite{had2}. Among mathematicians Hadamard
regularization is a rather standard technique in order to deal with
singular differential and integral equations with boundary conditions, both
analytically and numerically (for a sample of references see
\cite{had3}). Indeed, Hadamard regularization is a well-established
procedure in order to give sense to infinite integrals
\cite{eehad1}. Hadamard convergence is also one of the cornerstones
in the rigorous formulation of quantum field theory through micro-localization,
considered by specialists to be the most important step towards the
understanding of linear partial differential equations since the invention of distributions
(for a beautiful, updated treatment of Hadamard's regularization see
\cite{melr1}). In  \cite{eehad1} Hadamard regularization was invoked
in order to fill the gap between the infinities appearing in the quantum field theory
renormalized results and the finite values obtained in the
literature with other procedures. It was seen that the finite
results derived using Hadamard's regularization coincide with the
values obtained using the more classical and less rigorous methods
in the literature on the Casimir effect. Moreover, Hadamard's
prescription is able to separate and identify the singularities as
physically meaningful cut-offs. Although the strict significance of
this additional regularization is still not well understood, the
fact that it is able to bridge the two approaches is already
remarkable. In the following section we present a case in a much
related situation which can also serve as an example of the
regularization issue in the Deutsch-Candelas fashion: we will
clearly prove the advantages of using a prescription that, even if
mathematical in nature, is very well adapted to proposed laboratory
experiments.

\section{The Dynamical Casimir Effect (Davies-Fulling)}
The Davies-Fulling model \cite{fd76,fd77} describes the creation of
massless particles by a moving perfect mirror following a prescribed
trajectory. This phenomenon is also termed as the {\it dynamical
Casimir effect}. Moving mirrors modify the structure of the quantum
vacuum, what manifests in the creation and annihilation of
particles. Once the mirrors return to rest, a number of the produced
particles will generically still remain and can be interpreted as
radiated particles. This flux has been calculated in the past in
several situations by using different methods, as averaging over
fast oscillations \cite{dk96,jjps97}, by multiple scale analysis
\cite{cdm02}, with the rotating wave approximation \cite{sps02},
with numerical techniques \cite{ru05}, and others \cite{bmoo}. Here
we will be interested in the production of the particles and in
their possible energy values while the mirrors are in movement. This
is in no way a simple issue and a number of problems have
recurrently appeared in the literature when trying to deal with it.
To start, it is in this case far from clear which is the appropriate
regularization to use. Different authors tend to employ different
prescriptions, forgetting sometimes about the need to carry out a
proper (physical) renormalization procedure, as was also the case in
the other situations described in the preceding section. Thus, it
turns out that ordinarily, in the case of a single, perfectly
reflecting mirror, the number of produced particles as well as their
energies diverge all the time while the mirrors move. Several
prescriptions have been used in order to obtain a well-defined
energy, however, for some trajectories this finite energy is {\it
not} a positive quantity and cannot clearly be identified with the
energy of the produced particles (see e.g. \cite{fd76}).

The approach I will describe here is joint work with J. Haro
\cite{he2}, and relies on two very basic ingredients. First, proper
use of a Hamiltonian method and, second, the introduction of
partially transmitting mirrors, which become transparent to very
high frequencies. We have been able to prove in this way, both that
the number of created particles remains finite and also that their
energies are always positive, for the whole trajectories
corresponding to the mirrors' displacement. We have also calculated
from first principles the radiation-reaction force that acts on the
mirrors owing to the emission and absorption of the particles, and
which is related with the field's energy through the ordinary energy
conservation law. As a consequence, the energy of the field at any
time $t$ is seen to equal, with the opposite sign, the work
performed by the reaction force up to this time $t$
\cite{fv82,be94}. Such force is usually split into two parts
\cite{e93,bc95}: a dissipative force whose work equals minus the
energy of the particles that remain \cite{fv82}, and a reactive
force, which vanishes when the mirrors return to rest. It can be
seen that the radiation-reaction force calculated from the
Hamiltonian approach for partially transmitting mirrors satisfies,
at all time during the mirrors' oscillation, the energy conservation
law and can naturally account for the creation of positive energy
particles. Also, the dissipative part obtained within this procedure
agrees with the one calculated by other methods, as using the
Heisenberg picture or other effective Hamiltonians (but those
methods have traditionally encountered problems with the reactive
part, which in general yields a non-positive energy that cannot be
considered as that of the particles created at any specific $t$).

\section{A Consistent Formulation: Semitransparent Mirrors}
One of the main ingredients of the method is to use partially
transmitting mirrors, which become transparent to very high
frequencies (this is given by an {\it analytic matrix}). The second
main ingredient is proper use of a {\it Hamiltonian method} and the
corresponding renormalization. We proved both that the number of
created particles is {\it finite} and that their energy is always
{\it positive,} for the whole trajectory during the mirrors'
displacement. The  radiation-reaction force acting on the mirrors
owing to emission-absorption of particles is related with the
field's energy through the ordinary energy conservation law: the
energy of the field at any time $t$ equals (with opposite sign) the
work performed by the reaction force up to this time $t$. Such force
is split into two parts a {\it dissipative} force whose work equals
minus the energy of the particles that remain and a {\it reactive}
force vanishing when the mirrors return to rest.  The dissipative
part obtained agrees with the corresponding one from other methods.
But those have problems with the {\it reactive} part, which in
general yields a {\it non-positive} energy (which is not our case).
To be noticed is that several proposal at an experimental
verification of the dynamical Casimir effect have been issued
recently. \vspace*{1mm}

\ni{\bf Some Details and Examples.} We use a Hamiltonian method for
a neutral Klein-Gordon field in a cavity $ \Omega_t$, with
boundaries moving at a certain speed  $ v << c$, $ \epsilon =v/c$
(of order $ 10^{-8}$ in the experimental proposal \cite{kbo06}).
Assume the boundary is at rest for time $ t\leq 0$ and returns to
its initial position at time $ T$. The Hamiltonian density is
conveniently obtained using the method in \cite{js96}. The
Lagrangian density of the field is \beq {\mathcal L}(t,{\bf
x})=\f{1}{2}\left[(\partial_t\phi)^2 -|\nabla_{\bf
x}\phi|^2\right],\quad \forall {\bf x}\in\Omega_t\subset
\mathbb{R}^n, \ \, \forall t\in \mathbb{R}. \eeq   Transform now the
moving boundary into a fixed one by the (non-conformal) change of
coordinates \beq  {\mathcal R}:(\bar{t},{\bf y})\rightarrow
(t(\bar{t},{\bf y}), {\bf x}(\bar{t},{\bf y}))=(\bar{t}, {\bf R}
(\bar{t},{\bf y})),   \eeq which converts $ \Omega_t$ into a fixed
domain $ \tilde\Omega$:
 $ (t(\bar{t},{\bf y}), {\bf x}(\bar{t},{\bf y}))$
$ ={\mathcal R}(\bar{t},{\bf y})=(\bar{t}, {\bf R}(\bar{t},{\bf
y}))$ (with $ \bar{t}$  the new time).

The Hamiltonian density is \beq  \hspace*{12mm} \tilde{{\mathcal
H}}(\bar{t},{\bf y})=\frac{1}{2} \left(\tilde{{\xi}}^2
+J|\nabla_{\bf x}\phi|^2\right)
+\tilde{{\xi}}\left(\partial_{\bar{t}}\tilde{{\phi}}-\sqrt{J}\partial_t\phi\right),
\eeq being $ \tilde{\phi}$ the field, $ \tilde{{\xi}}$ the conjugate
momentum, and $ J$ the Jacobian of the change $ d^3{\bf x}\equiv
Jd^3{\bf y}$.
 It turns out that
 \beq  \hspace{-5mm}{{\mathcal H}}(t,{\bf x})= {{\mathcal E}}(t,{\bf x})
+\xi(t,{\bf x})<\partial_s{\bf R}({\mathcal R}^{-1}(t,{\bf x}))
,{\bf\nabla}_{\bf x}{{\phi}}(t,{\bf x})>   \eeq
 \[  \hspace{25mm}+\f{1}{2} \left.\xi(t,{\bf x}) \phi(t,{\bf x})
\partial_s(\ln J)\right|_{{\mathcal R}^{-1}(t,{\bf x})}.   \]
  As a simple example, for a single mirror following the prescribed
trajectory  $ R(\bar{t},y)=y+\epsilon g(\bar{t})$, we explicitly get
 \beq \hspace*{10mm} {{\mathcal H}}(t, x)= {{\mathcal E}}(t,x) +\epsilon
\dot{g}(t)\xi(t,x)\partial_x{{\phi}}(t, x).  \eeq

\ni{\bf Case of a single, partially transmitting mirror.} In the
original Davis-Fulling model \cite{fd76}, the renormalized energy is
negative while the mirror moves: cannot be considered as the energy
of the produced particles at time $ t$  [cf.~paragraph after
Eq.~(4.5)]. An interpretation of this fact is that a perfectly
reflecting mirror is non-physical. One should consider, instead, a
{\it partially transmitting mirror,} transparent to high
frequencies, what is indeed a mathematical implementation of a
physical plate, continuing our discussion in the preceding section.

Consider the trajectory $(t,\epsilon g(t))$. When the mirror is at
rest, {\it scattering} is described by the matrix
 \begin{eqnarray}
S(\w)=\left(\begin{array}{cc}
{s}(\w)&{r}(\w)e^{-2i\w L}\\
{r}(\w)e^{2i\w L}&{s}(\w)\end{array}\right).
\end{eqnarray}
This  $ S$ matrix is taken to be ($ x=L$ position of the mirror):
(i)  Real in the temporal domain:  $ S(-\w)=S^*(\w)$. (ii) Causal: $
S(\w)$ is analytic for Im $ (\w)>0$. (iii) Unitary:
$S(\w)S^{\dagger}(\w)=$
 { Id}. (iv) The identity at high frequencies:  $
S(\w)\rightarrow$ { Id,} when $ |\w|\rightarrow \infty$, $ s(\w)$
and $ r(\w)$ being {\it meromorphic} (cut-off) functions:
 the material's {\it permitivity} and {\it resistivity},
 respectively.

The results obtained are rewarding, and we can clearly see the
origin of the divergence, in the perfect boundary conditions case,
and its simple cure obtained in the semitransparent mirror case. In
fact, in this Hamiltonian approach the obtained force is:
\begin{eqnarray} &&\hspace*{-15mm} \langle\hat{F}_{Ha}(t)\rangle
=-\f{\epsilon} {2\pi^2} \int_0^{\infty}\int_0^{\infty}\f{d\w d\w'\w
\w'} {\w+\w'} \ \Re\left[e^{-i(\w+\w')t}\
\widehat{\dot{g}\theta_t}(\w+\w')\right] \nn
\\&& \hspace*{30mm} \times
[|{r}(\w)+{r}^*(\w')|^2+|{s}(\w)-{s}^*(\w')|^2] +{\mathcal
O}(\epsilon^2),
\end{eqnarray}
 the integral {\it diverges} for a perfect mirror ($ r\equiv
-1$, $ s\equiv 0$, ideal case), but {\it nicely converges} for our
partially transmitting (physical) one, where $ r(\w) \to 0$ and $
s(\w) \to 1$ as $ \w \to \infty$.
  Energy conservation is fulfilled: the dynamical energy at
any time $ t$ equals, with the opposite sign, the work performed by
the reaction force up to that time $ t$ \beq
\langle\hat{E}(t)\rangle=-\epsilon\int_0^t\langle\hat{F}_{Ha}(\tau)
\rangle\dot{g}(\tau)d\tau .\eeq The case of two partially
transmitting mirrors is not so different. A similar, albeit more
involved analysis, can be carried out \cite{he2}. No basic
obstruction to envisaged to extend our procedure to  higher
dimensions and  fields of any kind.

\section{Cosmological Imprint of the Casimir Effect?}
Although we still seem far from having an idea of what a
full-fledged theory of quantum gravity will look like in the end,
semiclassical approaches to this issue led to the seminal idea,
first clearly stated by Ya.B.~Zeldovich in 1968 \cite{zeld1}, that
quantum vacuum fluctuations, as a form of energy, must `gravitate',
that is, must enter into the vacuum expectation value of the
stress-energy tensor $ \langle T_{\mu \nu} \rangle \equiv - {\cal E}
g_{\mu \nu} $ on the rhs of Einstein's equations \beq
R_{\mu\nu}-\frac{1}{2}g_{\mu\nu}R=-8\pi G(\tilde{T}_{\mu\nu}-{\cal
E}g_{\mu\nu}). \eeq There may be subtleties in this argumentation,
e.g. those pertaining to the following question: do quantum vacuum
fluctuations fulfill the equivalence principle of general relativity? This seems to
have been settled down very recently \cite{eqpri1}, but there are
still contradictory answers in the literature \cite{eqpri2}. This
will affect {\it cosmology}, since  $ \tilde T_{\mu \nu}$
excitations above the vacuum are in fact equivalent, in a given time
slice, to a {\it cosmological constant}  $ \Lambda =8\pi G{\cal E}$.

 Recent observations yield the value \cite{sdsscol1}
 \beq \Lambda_{\mbox{\small obs}}  \ = \ (2.14\pm 0.13 \times
10^{-3}\ \hbox{eV})^{4} \quad \sim \ 4.32 \times 10^{-9}\
\hbox{erg/cm}^3. \eeq As we said, the cosmological constant gets contributions from
zero point fluctuations  \cite{zeld1} \beq  E_0 \ = \ \frac{\hbar\,
c}{2} \sum_n \omega_n, \qquad \omega = k^2 + m^2/\hbar^2, \ \ k =2
\pi /\Lambda. \eeq  But evaluating in a box and putting a cut-off at
maximum $k_{max}$ corresponding to reliable quantum field theory physics (e.g., the
Planck energy), one immediately gets the very huge number \beq \rho
\ \sim \ \frac{\hbar \, k_{\mbox{\footnotesize Planck}}^4}{16 \pi^2}
\ \sim \ 10^{123} \rho_{\mbox{\small obs}}.\eeq This is possibly the
largest discrepancy between theory and observation ever encountered
in physics.

Assuming one will be able to prove that the ground value of the cc
is {\it zero}, we will be left with this {\it incremental} value
coming from the topology or boundary conditions. This sort of two-step approach to
the cosmological constant is becoming very popular recently as the most accessible way
to try to solve this extremely difficult issue \cite{ccr2}. We have
then to see, using different examples, if this value acquires the
correct order of magnitude
---corresponding to the one coming from the observed acceleration in
the expansion of our universe--- under some reasonable conditions.
The idea is to involve the {\it global} topology of the universe
\cite{ct1}, in connection with the possibility that a faint scalar
field pervading the universe could exist. Fields of this kind are
ubiquitous in inflationary models, quintessence theories, and the
like. Also, given the fact that the universe expands, it is
plausible that the dynamical Casimir effect should play a role in
this discussion. Actually, one does not pretend in this way to solve
the old problem of the cosmological constant, not even to contribute significantly to
its understanding, but just to present simple and usual models which
show that the right order of magnitude of (some contributions to)
$\rho_V$ which lie in the precise range deduced from the
astrophysical observations may be not difficult to get. In different
words, we only address here the `second stage' of what has been
termed by Weinberg \cite{wei2} the {\it new} cosmological constant problem.
It should be mentioned however, in this context, that there are some authors
saying that the old cosmological constant problem is in fact trivial 
(see, e.g., \cite{volovik}). That in any
emergent gravity theory the natural value of the energy of the perfect
non-perturbed vacuum (and thus of the cosmological constant)
is exactly zero. The small, non-zero value that we see just
comes from perturbations of the vacuum, due to the expansion of the Universe,
gravitating matter, and effects of boundaries as discussed below. \ms

\ni{\bf Simple model with large and small compactified dimensions.}
We assumed the existence of a scalar field  extending through the
universe and calculated the contribution to the cosmological constant from the Casimir
energy density of this field, for some typical boundary conditions.
Ultraviolet contributions must be safely set to zero by some
mechanism of a fundamental theory. Another hypothesis is the
existence of both large and small dimensions (the total number of
large spatial coordinates being always three), some of which  may be
compactified, so that the global topology of the universe plays an
important role. There is a quite extensive literature both in the
subject of what is the global topology of spatial sections of the
universe \cite{ct1} and also on the issue of the possible
contribution of the Casimir effect as a source of some sort of
cosmic energy, as in the case of the creation of a neutron star
\cite{sokol1}. There are arguments that favor different topologies,
as a compact hyperbolic manifold for the spatial section, what would
have clear observational consequences \cite{css-mfo}. Other
interesting work along these lines was reported in \cite{elif12} and
related ideas have been discussed very recently in \cite{banks1}.
However, we differ from all those in that we put emphasis just in
obtaining the right order of magnitude for the effect. At the
present level it has no sense yet to consider more specifications
concerning the nature of the field, the different models for the
topology of the universe, and the different boundary conditions possible, with its
effect on the sign of the force too. This is left to future
analysis. From previous results \cite{zb12k} we know that the range
of orders of magnitude of the vacuum energy density for the most
common possibilities is not so widespread, and may only differ by at
most a couple of digits. This allows us, both for the sake of
simplicity and universality, to deal with two simple situations,
corresponding to a scalar field with periodic boundary conditions or spherically
compactified. As explained in \cite{eenc}, most cases with usual boundary conditions
reduce to those, from a mathematical viewpoint.

For lack of space we will not describe these models in detail here
(this has been done elsewhere \cite{eeqfext05}). Suffice to say that
it can be proven that the contribution of the vacuum energy of a
small-mass scalar field, conformally coupled to gravity, and coming
from the compactification of some small (2 or 3) and some large (1
or 2) dimensions ---with compactification radii of the order of 10
to 1000 the Planck length in the first case and of the order of the
present radius of the universe, in the second--- lead to values that
compare well with observational data, in order of magnitude, with
the exception of the {\it sign} ---which turns out to be opposite to
the one needed to explain negative pressure. To deal with this
crucial issue, we consider the two following classes of models.\ms

\ni{\bf Braneworld models.} Braneworld theories may hopefully solve
both the hierarchy problem and the cosmological constant problem. The bulk Casimir
effect can play an important role in the construction (radion
stabilization) of braneworlds. We have calculated the bulk Casimir
effect (effective potential)  for conformal and for  massive scalar
fields \cite{enoo2003}. The bulk is a 5-dim AdS or dS space, with 2
(or 1) 4-dim dS branes (our universe). The results obtained are
quite consistent with observational data.\ms

\ni{\bf Supergraviton theories} We have also computed the effective
potential for some multi-graviton models with supersymmetry
\cite{ss1}. In one case, the bulk is a flat manifold with the torus
topology $\mathbb{R} \times \mathbb{T}^3$, and it can be shown that
the induced cosmological constant can  be rendered {\it positive} due to topological
contributions \cite{cezplb05}. Previously, the case of
$\mathbb{R}^4$ had been considered. In the multi-graviton model the
induced cosmological constant can indeed be positive, but only if the number of massive
gravitons is sufficiently large, what is not easy to fit in a
natural way. In the supersymmetric case, however, the cosmological constant turns out
to be positive just by imposing anti-periodic boundary condition in the fermionic
sector. An essential issue in our model is to allow for
non-nearest-neighbor couplings.

For the torus topology we have got the topological contributions to
the effective potential to have always a fixed sign, which depends
on the boundary condition one imposes.  They are negative for periodic fields, and
positive for anti-periodic ones. But topology provides then a
mechanism which, in a natural way, permits to have a positive cosmological constant in
the multi-supergravity model with anti-periodic fermions. The value
of the cosmological constant is regulated by the corresponding size of the torus. We
can most naturally use the minimum number, $N = 3$, of copies of
bosons and fermions, and show that ---as in the first, much more
simple example, but now with the right sign!--- within our model the
observational values for the cosmological constant can be approximately matched, by
making quite reasonable adjustments of the parameters involved. As a
byproduct, the results that we have obtained \cite{cezplb05} might
also be relevant in the study of electroweak symmetry breaking in
models with similar type of couplings, for the deconstruction issue.
\medskip

\noindent{\bf Acknowledgments.} Based on work done while on leave at
Department of Physics and Astronomy, Dartmouth College, 6127 Wilder
Laboratory, Hanover, NH 03755, USA; in part in collaboration with G.
Cognola, J. Haro, S. Nojiri, S.D. Odintsov and S. Zerbini. The
financial support of DGICYT (Spain), project FIS2006-02842, and of
AGAUR (Generalitat de Catalunya), project 2007BE-1003 and contract
2005SGR-00790, is gratefully acknowledged.

\end{document}